# AI-Accelerated Brute Force Cryptanalysis

## *Trivial Ciphertexts Vulnerable*

Gideon Samid
Electrical, Computer and System Engineering
Computer and Data Sciences
Case Western Reserve University, Cleveland, OH
Gideon.Samid@CASE.edu

*Abstract: Modern cryptography is hinged on "not learning from mistakes": trying numerous wrong keys, should not help one identify the right key. Indeed, it worked -- until recently when the surprising power of AI to see pattern in apparent randomness has turned the 'wrong plaintexts' generated by the 'wrong key' into productive inferential input. Crunching through these random-looking plaintext candidates AI can de-flatten the probability curve over the remaining key space. The more spiked this curve, the faster the ciphertext is defeated. This new attack vector demands a thorough review of our cryptographic security posture. NIST PQC is not immunized against AI-Accelerated Brute Force attack. Defense is rooted in non-trivial ciphertexts, in unilateral randomness, and in variable key size. This points to a new security class: Pattern Devoid Cryptography which is to be added into the toolbox used by the cyber security community.*

# 1.0 Introduction

Modern cryptography is characterized by "committed ciphertexts". Namely, given an arbitrary ciphertext, C, generated by an encryption algorithm Enc from a plain language plaintext P, using a symmetric key K, there is no other {P', K'} pair where plaintext, P' ≠ P and key K' ≠ K such that the Enc matching decryption, Dec, will decrypt C into a plain language P'

$$\forall C \mid C = Enc(P,K)\ \neg\ \exists\ P' \neq P,\ \neg\ \exists\ K \neq K' \mid P' = Dec(C, K')$$

where:

$$P = Dec(Enc(P,K))$$

In simple terms our industry standard is limited to "committed ciphertexts" -- which are committed to the plaintext and the key that generated them. This mathematical commitment implies that with sufficient math wisdom and sufficient computational power a cryptanalyst will extract {P,K} from C. The entire science of cryptography is built on the premise that no attacker will be able to extract the {P,K} pair in a timely manner. Namely the efforts to extract "P,K" (To breach C) will become successful only after the secrecy of the plaintext P is void. Or say, the ciphers we use today rely on the assumption that no timely breach of their exposed ciphertext will occur.

Until recently there were four strategies for cryptanalysis:

1. Brute Force

2. Cipher Compromise

3. Weak Keys

4. Implementation Vulnerability

Brute Force -- trying all the keys in the key space until the right one is found -- is guaranteed to succeed. The defense was based on a credible estimation of the adversarial computation power, and adjusting the size of the key space to prevent a timely breach.

Cipher Compromise - math insight -- requires for the attacker to use greater mathematical talent than the defender. Very smart defenders, like the US National Institute of Science and Technology, NIST, are pretty confident that if their team does not see a mathematical breach, neither does the attacker.

Weak Keys -- are a subclass of keys in the key space, for which a compromise is found. If such weak keys represent x% of the key space then x% of the traffic is compromised. Here too, smart defenders assume that if there are any weak keys, they will be spotted by the defender first and would be avoided. [1].

Implementation vulnerability is also a question of smart v. smart and defenders are quite confident that they can cover all the angles.

Quite recently a new cryptanalytic strategy came forth, and it changes the balance between cryptography and cryptanalysis: Artificial Intelligence, AI .

To understand this new strategy, we need to remind ourselves of a fundamental attribute of a good cipher: *the avalanche effect*. It is this effect that AI attacks and thereby makes all common ciphers today vulnerable.

## 1.1 The Avalanche Effect

The defense against brute force attack is hinged on the assumption that to begin with an attacker facing a key space comprising $n$ possible keys, will have to associate each of those keys with a probability $1/n$ to be the key used to build the captured ciphertext. After testing $m < n$ keys, the chance for each of the remaining $(n-m)$ keys to be the right key is $1/(n-m)$. If this assumption holds and the defender has a good estimate of the attacker computational power than the security of the ciphertext can be credibly appraised.

This assumption is known as the *'flat key space assumption'*. It means that no matter how little from the key space remained unchecked (how large is *m*), the remaining *(n-m)* keys will show a flat probability curve at a *1/(n-m)* level.

In order to substantiate this assumption it is necessary to ensure that the 'wrong plaintext' decrypted from 'the wrong key' does not provide the attacker with any useful information with which to de-flatten the probability curve. This requirement was well recognized in DES which was designed to maximize the fact that change in one key bit will 'avalanche' throughout the ciphertext. Every serious cipher ever since is satisfying this requirement. Indeed, decrypting a ciphertext with a key with a small Hamming distance from the proper key will generate a plaintext candidate that is vastly different from the proper plaintext. Decryption with keys other than the proper one result in randomized plaintext which does not indicate to the cryptanalyst that the true key is very close to the one used.

While all major ciphers ensure the avalanche effect, the apparent randomness is only apparent. Von Neumann famously said that anyone generating randomness from algorithms, understands neither randomness, nor algorithms. There is no true randomness in the avalanche effect. The underlying pattern is simply buried too deep for classic cryptanalysis to detect.

Using the wrong key, K' over C, is resulting in the wrong plaintext, P'. P' has a Hamming distance, h-=Hamm(P,P') relative to P. Hence one needs to flip h bits in P' one by one, to get a series of plaintext of h options: P'....P before switching P' to P. This plaintext series appears random to the 'naked eye' but random it is not. It is derivable from the cipher, C and K. Existing pattern that is too complex for human identification is the perfect input for AI pattern extraction.

## 1.2 The Avalanche Effect as an AI Vulnerability Feature

Modern AI sees patterns where non-AI tools see strict randomness. The neural networks inferential mechanism is beyond human tracking, yet with a remarkable record of success. The greatest success was achieved with respect to supervised AI where a definite target function is optimized by analyzing relevant data.

Given a mainstream cipher algorithm CA, it is common for both the encryption and the decryption algorithms to be in the public domain. This is known as The Kerckhoffs's principle. Accordingly, a cryptanalyst capturing a ciphertext, C, which was encrypted from a plaintext $P_0$ using key $K_0$, will decrypt it with any arbitrary number of keys: $K_1$, $K_2$, ....$K_m$, then list m respective plaintext candidates $P_1$ $P_2$, ..... $P_m$.

The accumulated data of *m* keys and m corresponding plaintexts represent data that appears random, but has no random data in it, it is computational complexity appearing as 'random'. It is thus a fertile ground for AI analysis for the purpose of 'spiking up' the otherwise flat probability curve for the yet untried keys.

The more keys one tries (the larger *m*), the more data to analyze and the more de-flattened the probability curve. The less-flat the curve the fewer keys to be tried before the right key is spotted.

From the security point of view, it is unknown how advanced the adversarial AI algorithms are, and how fast the resultant cryptanalysis. Unlike a mathematical breach of a cipher, this AI assisted cryptanalysis. (AIACA) is getting more effective with time. It is hard for the security community to assess how effective is the adversarial AI.

## 1.3 Literature Survey

Ronald Rivest, [17], declared "Machine learning and cryptanalysis can be viewed as Sister fields, since they share many of the same notions and concerns." Ever since this observation has been challenged because cryptography deals with exact values, while AI is most powerful by dealing with progressive approximations, and measured similarities.

It is only lately that the literature has addressed the question of AI as a new worrisome cryptanalytic tool. Some trace the interest to Gohr's article [20] which addressed a very specific case, ( Speck32/64). Gohr concluded that "In the setting considered, this works reasonably well". The neural networks needed only a few minutes to be trained. This led Gohr to insist that AI may develop to become a very powerful cryptanalysis agent. Gohr asserts: "This paper is the first to show that neural networks can be used to produce attacks quite competitive to the published state of the art against a round-reduced version of a modern block cipher."

Several researchers have applied AI techniques for other limited situations (DES, RC4) [21, 22, 23, 24]. None reported an alarming breakthrough. More recent publications, [19], report limited attempts to unleash neural networks on a cryptanalytic challenge, with no impressive results. These attempts may be judged as too narrow.

All in all the literature reflects a timid approach to the proposition that advanced AI is rising to become a front line serious threat vector to mainstay cryptography. By comparison, Peter Shor published his paper on the threat of quantum computers [25] in the early 90's. It took almost a quarter of a century for the cryptologic community to arm itself against this threat. Nonetheless AI today is much more advanced than quantum computing was in the 90s. The AI threat vector may present itself much faster.

# 2.0 Methodology

We describe two methodologies: (i) General Intelligence, AI, an (ii) Supervised, goal-seeking AI. We also (iii) present a bird's view of the avalanche challenge. The source data may be: (i) brute force data only, or (ii) brute force data plus cipher analysis data.

Using General Intelligence AI one will identify (i) a target plaintext as plaintext string that reads like a proper plain language statement -- complying with the expected redundancy of the language in point (2.3 bit/letter for English). (ii) the description of the attacked cipher algorithm, (iii) the key space, and optionally (iv) examples of prospective plaintexts, representing the environment where the encryption takes place. Alternatively these prospective plaintexts will be issued by the AI which is familiar with the situation. The goal for the AI is to find a key that will decrypt the subject ciphertext to a plausible plaintext.

Supervised goal-seeking AI is less sophisticated than general intelligence AI, but as of now, it is more advanced and more powerful.

Let $\delta(K_0, K_i)$ be a well-defined 'distance' measure between $K_0$, the key used to encrypt the secret plaintext $P_0$ to the ciphertext C, and any other key, $K_i$ in the key space. The self-distance is set to zero: $\delta(X,X)=0$. The most common formula for δ is the Hamming distance.

A general algorithm. ALG, such that: $P_0 = ALG(C, K_0)$ is expected to hold:

***If*** $\boldsymbol{\delta(K_0, K_i) < \delta(K_0, K_j)}$

then $\delta(P_0, P_i) < \delta(P_0, P_j)$

In a cipher this 'distance' matching is disrupted to interfere with a process of gradually inching towards the right key. In a good cipher a small change in the key generates a large change in the computed plaintext (the avalanche effect). This large change appears random, but it is pattern bearing.

A cryptanalyst computing *m* plaintext candidates $P_1, P_2, ...P_m$ from *m* key candidates: $K_1, K_2, .... K_m$ will try to deduce the best next key, $K_{m+1}$ to apply, so that after a minimum of tried keys the right one, $K_0$, will be found.

The cryptanalyst will gain more insight into the cipher behavior by using cipher analytics data.

# 2.1 Cipher Analysis Data

Given a particular cipher algorithm CA, one could use an arbitrary $P_0$ and arbitrary $K_0$ to generate a ciphertext $C_0$. Then one could select m arbitrary keys, $K_1, K_2, ...K_m$, apply them to $C_0$ and generate *m* plaintext candidates $P_1, P_2,.... P_m$.

One would then select *d* distance measurements for same size binary strings, x, y:

$\boldsymbol{\delta_1(X,Y), \delta_2(X,Y)..... \delta_d(X,Y)}$

and apply these *d* distance metrics to compute:

$\boldsymbol{\delta_j( K_i, K_0)}$***..... for i=1,2,....m and j=1,2,...d***

as well as:

$\boldsymbol{\delta_j( P_i, P_0)}$***..... for i=1,2,....m and j=1,2,...d***

Next one will select $\delta_k$ to map the $K_1, K_2, ...K_m$ on the x-axis, each key represented by its distance measure from $K_0$, $k \in \{1,2,...d\}$, and also select $\delta_p$ to map the $P_1, P_2, ...P_m$ on the y-axis, each plaintext candidate represented by its distance measure from $P_0$, $p \in \{1,2,...d\}$. These two selections ($\delta_k$, $\delta_p$) project m points on the surface x-y. These *m* points exhibit a pattern, that points to other keys, $K_{m+1}$, $K_{m+2}$ which are closer to the origin of the surface x-y ($K_0$, $P_0$).

This pattern will be used by the cryptanalyst after trying m keys $K_1$, $K_2$, ...$K_m$ over a captured ciphertext C, pointing to key candidates $K_{m+1}$, $K_{m+2}$...$K_{m+t}$ that are associated with probability $Pr_{m+t} > 1/(n-m-t)$ which amounts to de-flattening of the probability curve for the remaining *n-m* keys.

This "spiked" probability curve is open ended. The less flat it is the shorter the time for compromising the ciphertext C.

### 2.1.1 Distance Metrics

The common way to compute distance between two same size binary strings is the Hamming Distance: a count of bits of opposite identity. Obviously, any string has a zero distance with itself. There are up to

$$n! /(n/2)!^2$$

variations for n-bits strings. And they all fit into a Hamming distance range of 0-n. Namely there are many strings *s* that all have a particular Hamming distance *h* (the number is higher, the closer h is to n/2).

Early indications show greater utility with other distance metrics, especially the *3-summary distance.*

The **3-summary procedure** takes a binary string *s* comprising *n* bits and parcels it out three bits at a time. Each 3-bit sequence is summarized to its majority state: 1 if the summarized bits comprise two or three "1"s and "0", otherwise. Special rule applies to the leftover bits that no longer divide by three.

For example: s = 001 000 110 010 101 111 001 100, 011 is summarized to s' = 101011001

Applying iteratively:

s '= 101 011 001 is summarized to s'' = 110

Two strings *x*, and *y* of *n* bits both, may become identical after *t* rounds of 3-summary operations. The larger the value of *t*, the greater the distance between x and y.

Instead of 3-summary one can apply *q*-summary, where *q* is any odd number.

Distance metrics should be taken generously, to allow AI neural networks to find ever more efficient accelerators for brute-force cryptanalysis. There are in theory infinite ways to formally define a distance between binary strings. The literature is quite reach: (i) Levenshtein Distance, (ii) Jaccard Distance, (iii) Cosine Distance, (iv) Euclidean Distance, (v) Manhattan distance, (vi) longest common sub-sequence. [9]

# 2.2 Direct AI accelerated Brute Force ($AI^2$)

Describing here a method that was tried on a limited basis and has shown promise ($AI^2$). AI is used here two ways: (i) listing plaintext candidates, (ii) searching for pattern in a progressive series.

**Situation:** A cryptanalyst captures a series of ciphertexts sent from a given transmitter to a given recipient. These ciphertexts reflect the security situation which is defined by environmental parameters known to the cryptanalyst. On the basis of these parameters the AI is tasked to issue a series of plausible plaintext candidates for each captured ciphertext. Issuing plausible plaintext is a hard task. [12].

Let the ciphertext series be $C_1, C_2, ....C_c$. Based on the security situation in general and based on the order, frequency, length, etc. of these ciphertexts, AI is tasked with issuing a list of plausible plaintext candidates for the c ciphertexts. Plaintext candidate $\pi_{ij}$ will be plaintext candidate j for ciphertext i.

$$C_i: \{\pi_{i1}, \pi_{i2}, ..... \}$$

Let $C_q$ ($q \in 1...c$) be the ciphertext for which the plaintext candidates are evaluated to be most credible. [12]

Next the cryptanalyst will select $t$ keys $K_1, K_2, ...K_t$ and use them to decrypt $C_q$: $P'_1, P'_2,..... P'_t$ plaintext candidates respectively:

$$P'_i = Dec(C_q, K_i). .... \text{ for } i=1 \text{ to } i=t$$

Selecting a distance metric $\delta_k$, AI will compute the distances between each of the $t$ plaintext candidates with each of the plausible plaintexts $\pi_{q1}, \pi_{q2},....$ .

Next one selects for each $P'_i$ (i=1,2,..t.) the corresponding $\pi_{qj}$ for j=1,2,3,... for which

$$\delta_{ki} = \delta_k(P'_i, \pi_{qj})_{min}. \text{ for } i=1,2,...t \text{ and } j=1,2,...$$

Next the cryptanalyst will rank order the $t$ $P'_i$ according to the minimum distance they exhibit with their respective plausible options. Once this order has been established then the $t$ keys $K_1, K_2, ...K_t$ are ordered respectively. The $t$ keys are now ordered in an order that tracks how close these keys are in terms of producing plaintext candidates that are close as possible to the plausible plaintexts. The $t$ keys are then order-marked as:

$$\Omega = K^1, K^2, .....K^t$$

Here we come to the second inferential AI task: to examine the ordered series $\Omega$ with an attempt to unveil a pattern in $\Omega$ that would be consistent with the series and further suggest the $t'$ keys to be used in a repetition of the above.

If as repetitions are performed the computed minimum distances get smaller and smaller then this procedure should be continued until the right key is stumbled on, which should be much earlier than with plain blind brute force.

If the distance values do not get smaller and smaller then one switches to another distance metric [12].

If the computed distances don't become smaller fast enough then this entire procedure is run over another choice of the captured ciphertexts.

This extensive AI procedure is designed to extract a series of keys that are getting closer and closer to the right key, homing in, to hit this right key much earlier than at half key space.

## 2.3 Reverse Avalanche

The avalanche effect is a hard requirement of any serious cipher: flipping a single bit in the key, should 'avalanche' into a 'completely different' ciphertext. Hence a progressive series of such 'completely different' ciphertexts bears a hidden order: flipping one bit after another in the respective key. For an AI system that goes through extensive learning, running on many examples, any algorithmic confusion masqueraded as randomness will eventually be flushed out, as has been shown in various "black box" challenges, where highly complex functions were accurately extracted [26]. Combining this with the strategy to de-flatten the probability curve over the untested key space, points to a serious cryptanalytic tool.

Let a ciphertext C be decrypted with the proper key, $K_0$ to yield the proper plaintext $P_0$:

$$P_0 = Dec(C, K_0)$$

Let key $K_1 \neq K_0$ be a cryptanalytic key guess:

$$P_1 = Dec(C, K_1)$$

Obviously $P_1 \neq P_0$. Let the Hamming distance between $K_1$ and $K_0$ be h:

$$h = Hamm(K_0, K_1)$$

We mark $K_{oi}$ as a key string generated from $K_0$ by flipping *i* bits in $K_0$.

$$i = Hamm(K_0, K_{0i})$$

We may now write the series $K_{00}, K_{01}, K_{02},...K_{0h} = K_1$

Computing *h* plaintext candidates $P_{0i}$:

$$P_{0i} = Dec(C, K_{0i}) \text{.........} \textit{ for } i=0,1,2...h$$

The well-ordered series $P_{00}$, $P_{01}$, $P_{02}$,...$P_{0h}$ will appear random (the avalanche effect), but in fact this series projects on the K series: $K_{00}$, $K_{01}$, $K_{02}$,...$K_{0h}$ where each string is only one flipped bit different from its adjacent strings. The task before AI is to 'reverse the avalanche' appearance in the P series and extract the corresponding K series. Since everything except the key is in the open, AI has everything it needs to establish rules to apply to candidate keys in order to build a non-flat probability curve over them and shorten the time to hit the right key.

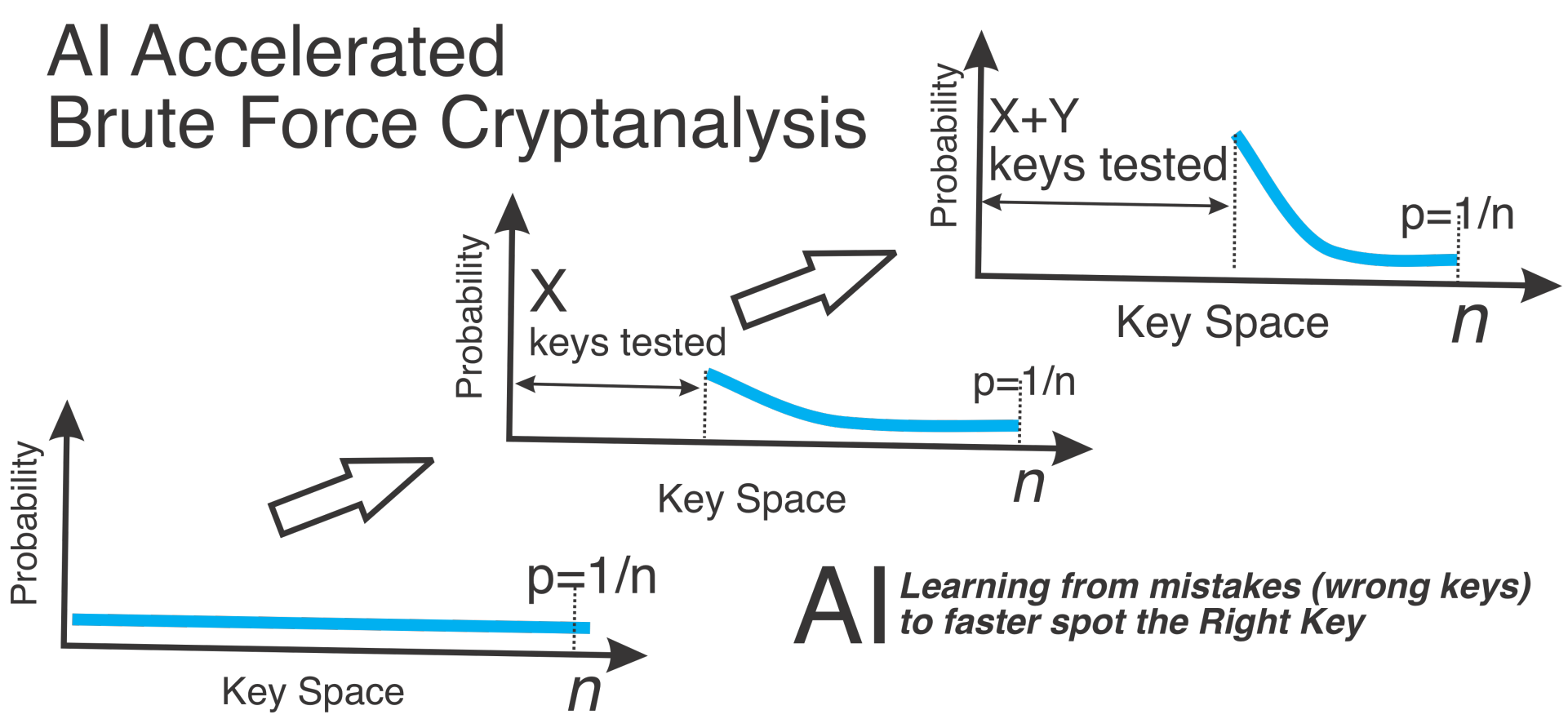

# 3.0 Defense

Cryptanalytic defense (cryptography) normally comes with the following categories:

1. sustainable mathematical complexity

2. flat probability, large enough key space

3. careful implementation.

AI Assisted Brute Force (AIA-Brute Force) attack is not defeated by mathematical complexity. It is not attacking implementation vulnerabilities, it attacks flat-probability curve over the key space. Defense, therefore should be hinged on a large enough key space to offset the spiked probability curve over the key space, but the defender would not know how 'spiked' the probability curve became, and to what extent is the defense effective.

Well planned defense would shatter the underlying assumptions of the AIA BF attack, which are:

1. Known Key Size

2. Key Durability

3. Trivial Ciphertext [15]

If the key size is unknown, then it is impossible to measure distance between it and a key candidate. The AIA-Brute Force attack will not work. Neither would the plain brute force attack work. If the key changes throughout the generation of the ciphertext, then also the AIA-BF does not work, nor does the plain brute-force.

Trivial ciphertext is a ciphertext where every bit represents content (is content-bearing) and is crucial in deciphering the ciphertext to its generating plaintext. A non-trivial ciphertext is one that contains a mix of content-bearing bits and content-devoid bits such that the intended reader will distinguish between these two categories, but the attacker won't. Non-trivial ciphertexts will be resistant against AIA-Brute Force attack. [15]

Claude Shannon defined the concept of "unicity distance" as the entropy of the key space divided by the redundancy of the language, D: UD = H(K)/D about 2.3 bits/letter in English, [3] showing that longer ciphertexts commit to their generating plaintext. This applies to trivial ciphertexts. There is no unicity distance for non-trivial ciphertexts [15].

While virtually all the mainstream ciphers today are of known key size, exhibit key durability, and generate trivial ciphertext, there is a pool of new ciphers that are based on unknown key size, use fast shifting keys and deploy non-trivial ciphertexts -- they are ready to withstand AIA Brute Force attack.

## 3.1 Randomness Defeats AI

AI extracts pattern from situations that any non-AI tool, including human intuition, has concluded to be 'purely random'. AI neural networks discern order that remained veiled before the best non-AI math tools. AI is therefore truly revolutionary. AI spots diseases, discovers new medicine, discerns archeological data, and as practiced by the author establishes AI Assisted Innovation [6] all beyond non-AI capability. Cryptography is no exception.

Yet, AI can be "poisoned", it can be defended from. In particular, randomness defeats AI onslaught. Randomness is a subtle notion. Any finite random series examined by AI will yield some expressed pattern that would be totally false, namely not a pattern that was hidden in the apparent randomness, rather AI hallucination.

Applied to cryptography randomness can be enhanced over the key, and over the ciphertext, creating a randomness-packed cipher that can withstand quantum attack, AI attack, and any yet unimagined pattern-seeking attacks on the captured ciphertext.

The shared key used by the communicators is random, however in all common ciphers the key is of known size and durable throughout the communication session and beyond. Extra randomness can be added by building a key of secret size, and using a randomized part thereto for every new session, even for every new statement, or with maximum variability, changing the key for every new letter -- totally voiding any AIA-BF attack.

A potent and robust defense is one that resorts to ciphers that allow for unilateral randomness to be injected by the transmitter such that the intended reader will identify it and ignore it but the AIA BF cryptanalyst will vainly attempt to spot pattern thereto.

It is noteworthy that NIST recently recognized the cryptographic power of unilateral randomness which is well used in its "Learning With Errors" Cryptography. [8].

A whole new class of ciphers exemplifies the above and serves well as AIA-BF defense. The class is known a Pattern Devoid Cryptography [2]

# 3.2 AIR-AI: AI resistant AI: AI Perfect Cipher

AI can be used to construct the perfect security cipher that resists all cryptanalytic attempts including AI cryptanalysis. Claude Shannon has proven [3] that the Vernam Cipher rightly implemented, offers perfect secrecy. He proved it by showing that knowledge of the captured ciphertext does not change the cryptanalyst outlook on the set of probable plaintexts. Namely before capturing the ciphertext, the cryptanalyst listed *n* plaintext candidates $P_1$, $P_2$, ...$P_n$ to be the plaintext encrypted and sent out as the public clear ciphertext. And after omnipotent hammering and squeezing of the ciphertext, the cryptanalyst remained with the same list of candidates. In other words, security is perfect if knowledge of the ciphertext does no impact the series of probable plaintext candidates.

Applying the same principle here let $P_1$, $P_2$, ....$P_n$ be the list of probable plaintext candidates. While $P_1$ is the plaintext that is actually sent to the intended recipient. The transmitter will choose *n* random keys $K_1$, $K_2$, .... $K_n$ then encrypt $P_i$ with $K_i$ to generate $C_i$. Next the transmitter will cast ciphertexts $C_2$, ..$C_3$, .... $C_n$ as noise relative to key $K_1$ and combine the (n-1) decoy strings with $C_1$ to build the combined ciphertext CC. The intended reader will ignore the decoys and decrypt CC to $P_1$ (using the shared key $K_1$). The attacker using AI, even privy to quantum computing , and being omni math-talented, at most will extract the *n* plaintext candidates they were listing before cryptanalyzing CC, and hence CC offers perfect security.

AI is used by the transmitter to generate plaintexts $P_2$, $P_3$,... $P_n$ from knowledge of $P_1$. [12]

# 3.3 Short Review of Common PDC Ciphers

Here is how **BitFlip** [4] works: given an alphabet A comprising *n* letters $a_1$, $a_2$, ... $a_n$, let every letter be associated with an arbitrary/random number of key bit strings, where each key string is of arbitrary/random size. These key strings are considered the cryptographic key. When Alice wishes to send Bob letter $a_i$ she randomly selects one of the key strings, associated with $a_i$, $k_{ij}$ and sends Bob a string $s_i$ which marks a Hamming distance h from $k_{ij}$:

$$h = Hamming(k_{ij}, s_i)$$

Note: The Hamming Distance can be replaced by any of the other distance metrics mentioned herein as well as others [9].

Then Alice checks the Hamming distance between $s_i$ and all the strings associated with the rest of the alphabet (this is the 'Confusion Test'). If any of these Hamming distances equals to h, then Alice builds another string $s'_i$ for which:

***h = Hamming($k_{ij}$, $s'_i$)***

and sends it out. Then she runs again the Confusion Test. If it fails, she repeats until the Confusion Test holds.

Bob will ignore all strings that fail the confusion test and read the letter $a_i$ once it is confusion free.

This simple routine is exercised letter after letter. Alice can also send Bob a string that shows no Hamming distance h from any key string -- Bob will ignore it too.

The BitFlip cipher is immunized against AIA Brute Force attack. The key is of unknown size, there is no complex math to short-cut, and the ciphertext is not trivial, it contains an unknown amount of noise.

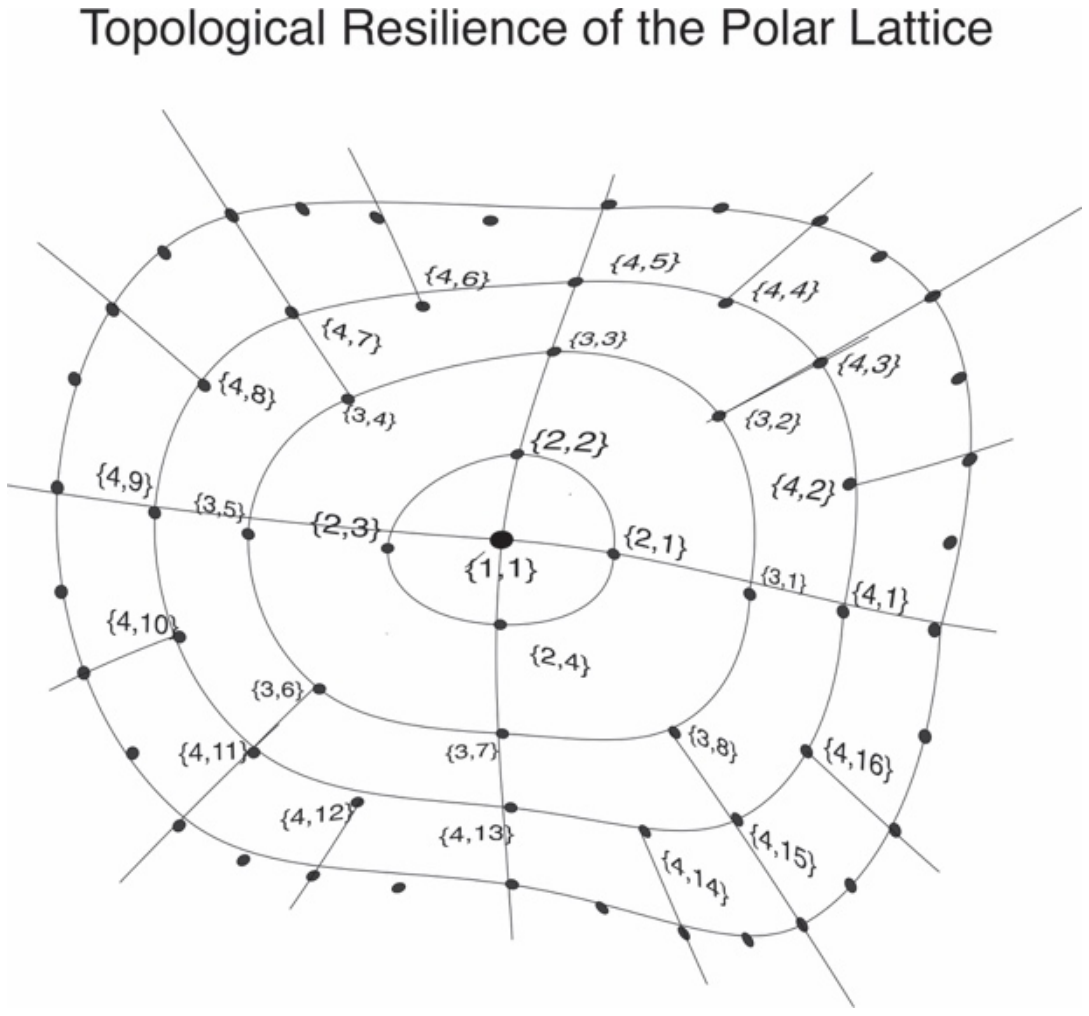

**Polar Lattice Cryptography** is another example for Pattern Devoid Ciphers that are immunized against AI Accelerated Brute Force attack [ 5 ]. Consider an arbitrary number of concentric circles with an arbitrary number of rays emanating from the shared center. Different rays extend to different circles. Every intersection between a circle and a ray is an addressable point. Each letter of the alphabet is associated with a starting point and a terminal point. When Alice wishes to communicate to Bob a given letter she marks an arbitrary path from the respective starting point to the respective terminal point. The path is defined by a series of points. Steps are taken: up/down (along the ray), right/left (along the circle) from one addressable point to the next. If the path leads from the starting point of letter $a_i$ to the terminal point of letter $a_i$, and there is no confusion with any other letter then Bob knows that Alice sent the letter $a_i$. Here too, the ciphertext includes open ended unilateral randomness that flows from Alice to Bob -- AIA-BF attack defeated.

# 4.0 AIAI

The thesis, tools and methodologies herein were developed by using Artificial Intelligence Assisted Innovation, AIAI, an innovation package initially developed in the PhD dissertation of the author at the Technion -- Israel Institute of Technology. [6]. So guided, an abstraction route was applied, pointing out that cryptography at its essence projects 'phony randomness': releasing mathematical constructs that impress their casual and malintent viewers as noise, randomness, pattern devoid, while delivering a clear pattern to their intended reader (the hidden plaintext). In the prevailing ciphers randomness is counterfeited through complexity, and hence a smarter mathematician will always prevail. This led to the different path to construct ciphers that would be inherently pattern devoid, project real randomness wrapping up the delivered secret. This was implemented through the use of unilateral randomness injected by the transmitter without pre coordination with the recipient. In other words: instead of hiding a secret in spurious randomness, hiding the secret in true randomness. It led to a series of ciphers, all abiding by the same principle, [7].

# 5.0 Outlook

"Learning from mistakes" is a fundamental strategy prevailing in the Darwinian evolutionary process that shaped the human brain. Today, AI is surprising even its designers, as to the power of learning from mistakes and approaching a stated goal. Modern cryptography, applying the 'avalanche effect' is bucking this trend, preventing a cryptanalyst from learning from their mistakes and approach a goal. This avalanche effect built up difficulties to 'learning from mistakes'. The current literature shows mild interest in this new threat vector. Several researchers report modest success. These first slow steps cast a misleading impression. In principle today's cryptography relies on complex mathematics. This is the stuff AI is good at cracking, even though nothing spectacular has been shown yet with respect to cryptography. There are plenty of examples [26] demonstrating how neural networks, Gaussian processes and symbolic regression retrieve mathematical complexities. AI is free to 'play' with any cipher, compare input and output to extract its hidden patterns. AI is extremely powerful with approximations and similarities and therefore, as presented here, the accelerated brute force strategy appears the most promising. And while not much academic interest has been demonstrated so far, it may well be that a great deal of action is hidden in the clouds of global geopolitics.

The impact of quantum computing on cryptography took a quarter of a century to be properly regarded. We should be faster with respect to the AI threat. We do have robust solutions. The general outlook for cryptography is to be moving away from trivial ciphertexts and known, fixed size keys. Future cryptography will contaminate' the ciphertext with unilateral randomness and deploy secret size keys which are randomly cut. Cryptography of the future appears likely to migrate from "Hidden Pattern Cryptography", HPC, today to "Pattern Devoid Cryptography", PDC [2] tomorrow.